\documentclass[multphys]{svmult}

\usepackage{makeidx}         
\usepackage{graphicx}        
\usepackage{multicol}        
\usepackage[bottom]{footmisc}
\usepackage{url}             

\makeindex             

\begin{document}

\title*{Validation of stellar population and kinematical analysis of galaxies}
\titlerunning{Kinematics and Stellar Populations of Galaxies: Validation}
%
%
\author{M. Koleva\inst{1,2}\and N. Bavouzet\inst{3}\and I. Chilingarian\inst{2,4}\and
P. Prugniel\inst{2,3}}
%
%

\institute{ Department of Astronomy, Sofia University, Bulgaria
\and CRAL -- Observatoire de Lyon, France 
\and GEPI -- Observatoire de Paris, France
\and Sternberg Astronomical Institute, Moscow, Russia}
%
%
\authorrunning{Koleva et al.}
\maketitle

\begin{abstract}
3D spectroscopy produces hundreds of spectra from
which maps of the characteristics of stellar populations (age-metallicity)
and internal kinematics of galaxies can be derived. We carried on 
simulations to assess the reliability of inversion methods and to define the
requirements for future observations. We quantify
the biases and show that to minimize the errors on the kinematics, age and 
metallicity (in a given observing time) the size of the spatial elements and
the spectral dispersion should be chosen to obtain an instrumental velocity 
dispersion comparable to the physical dispersion.

\end{abstract}

\section{Analyzing the data}
\label{sec:1}
Recently it became possible to derive simultaneously
internal kinematics and characteristics of the stellar population of galaxies
\cite{ocv05}. 

We use a simple parametric procedure \cite{chil05} to fit the moments of
the line-of-sight velocity distribution (LOSVD): v, $\sigma, ...$; and
parametrized star formation history (SFH) containing either single stellar
population (SSP) or several star bursts, using models computed with PEGASE.HR
\cite{leb04}. Before running the inversion procedure one needs to: (a)
determine variations of the spectrograph's line-spread function along
wavelength and over the field of the IFU; (b) inject this information into
template spectra.
We stress that 
the population parameters are constrained by the absorption lines and not by 
the shape of the continuum which may be affected
by internal extinction or calibration uncertainties.

\section{Validation of the population pixel fitting}
\label{sec:2}
The questions we address are: 
\begin{enumerate} 
\item Does the method return unbiased estimates of kinematics, age ($t$) and
metallicity ($Z$) at any signal to noise ratio (SNR)?
\item What are the degeneracies between different parameters?
\item What SNR and spectral resolution are required to get a given
precision?
\item Are various models of stellar populations consistent?
\end{enumerate}

\begin{figure}
  \centering
  \includegraphics[height=5cm]{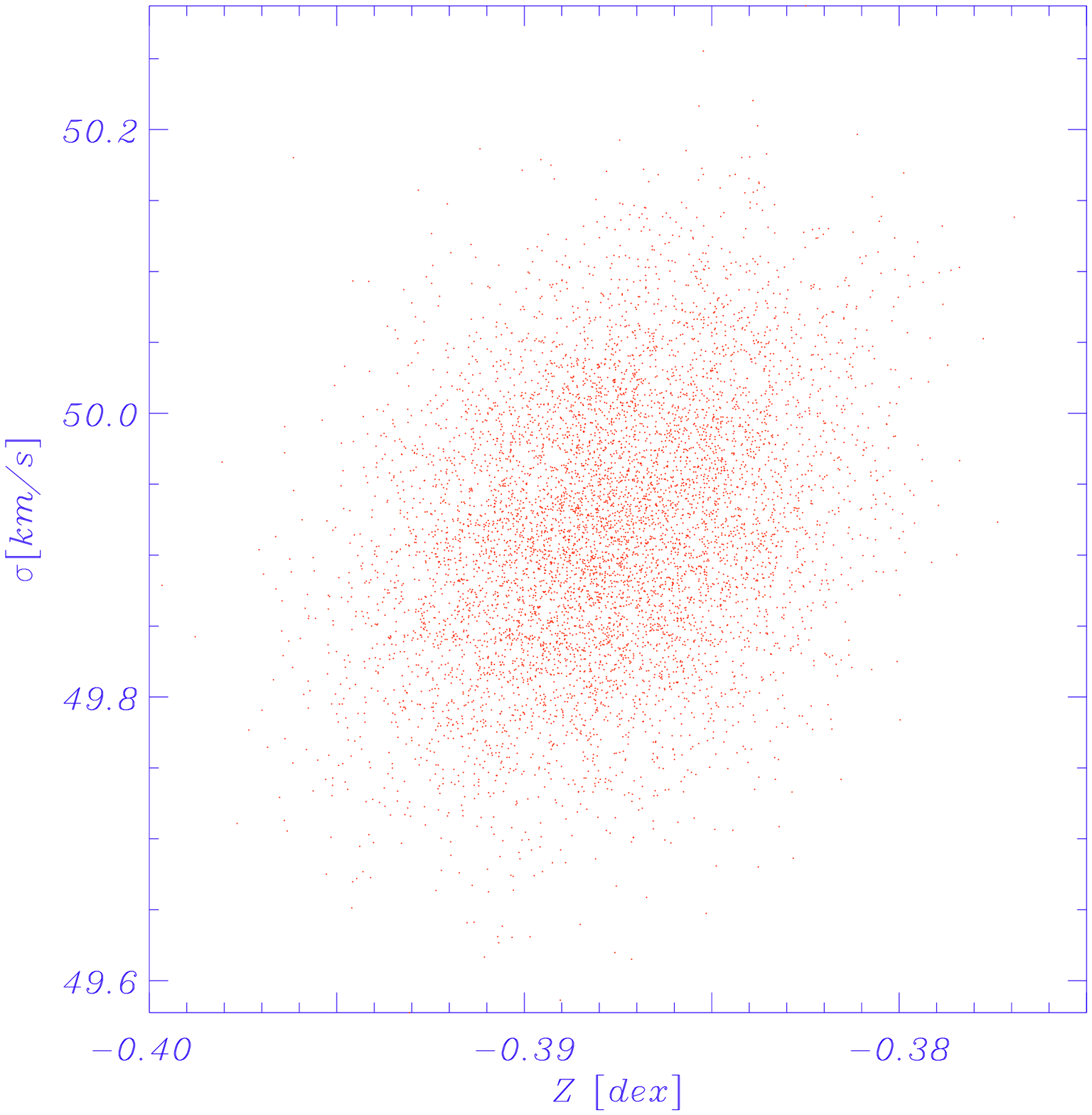}
  \includegraphics[height=5cm]{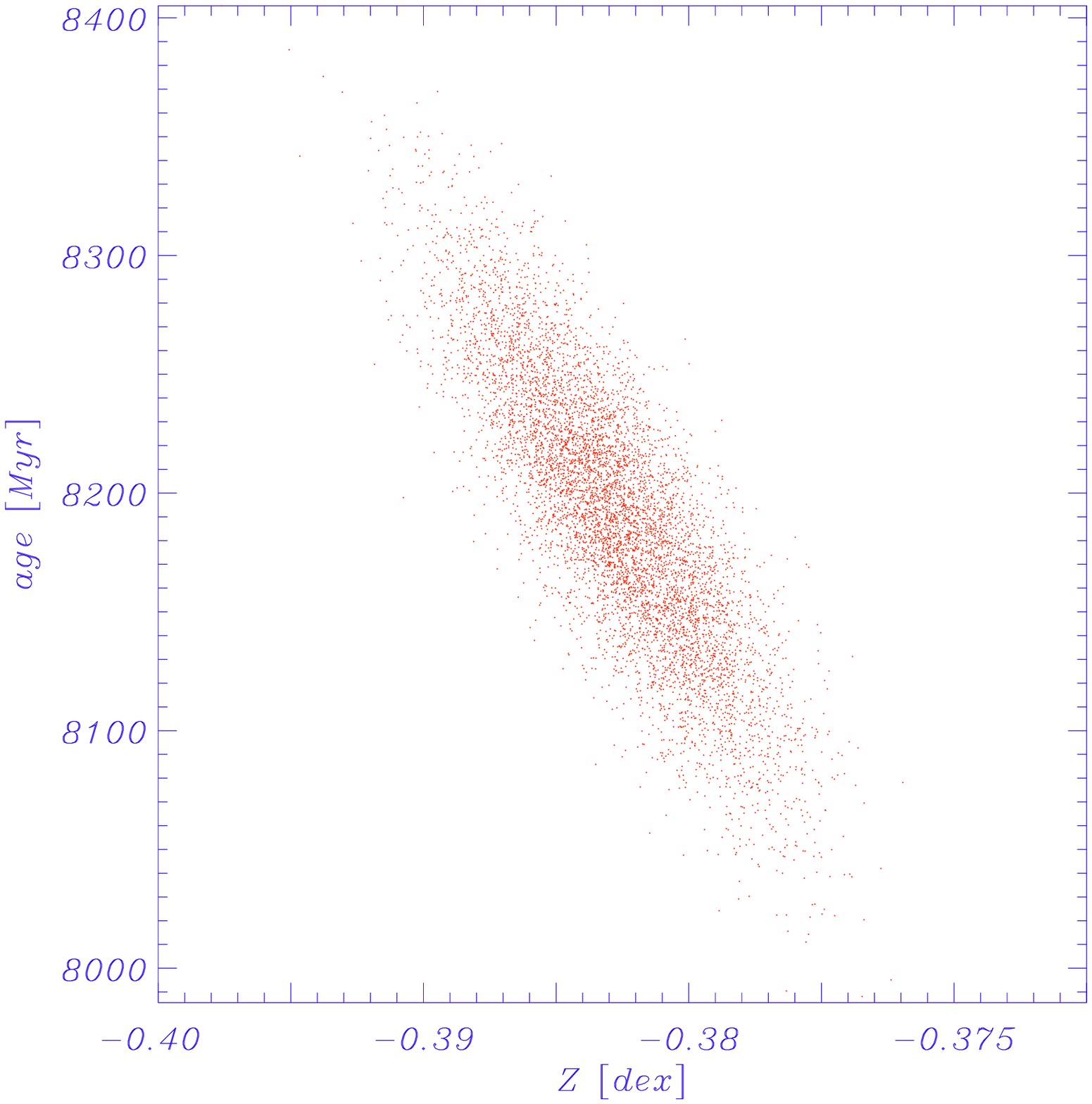}
  \caption{Metallicity-velocity dispersion and Age-metallicity  
    degeneracies. Monte-Carlo simulations with PEGASE.HR spectra,
    SNR = 100\,pix$^{-1}$, $\sigma$ = 50\,km\,s$^{-1}$}
  \label{fig:1}
\end{figure}

\emph{Biases and degeneracy.} We have performed extensive Monte-Carlo
simulations to fit SSP or SFH containing two bursts with SNR ranging from 1
to 500 pix$^{-1}$ (at R=10000; pix=0.2\AA). As long as the grid of models
is fine enough for performing good interpolation, our method is not biased
down to SNR=5 pix$^{-1}$. The main degeneracy is naturally between
age and metallicity, but Fig~\ref{fig:1} presents also the 
degeneracy between $\sigma$ and $Z$: a discrepancy toward higher $Z$
(sharp lines) is compensated by higher $\sigma$. This latter
degeneracy is the strongest coupling between the kinematical and population
parameters. An error of 1 dex on the metallicity results in an error 
of about $25 \%$ on the velocity dispersion. This can introduce a
significant systematics in the determination of the mass-to-light ratio
of galaxies. This effect is considerably reduced if an additive continuum is
included in the fit, but in this case constraints on the stellar populations are lost.

\begin{figure}
  \centering
  \includegraphics[height=15cm]{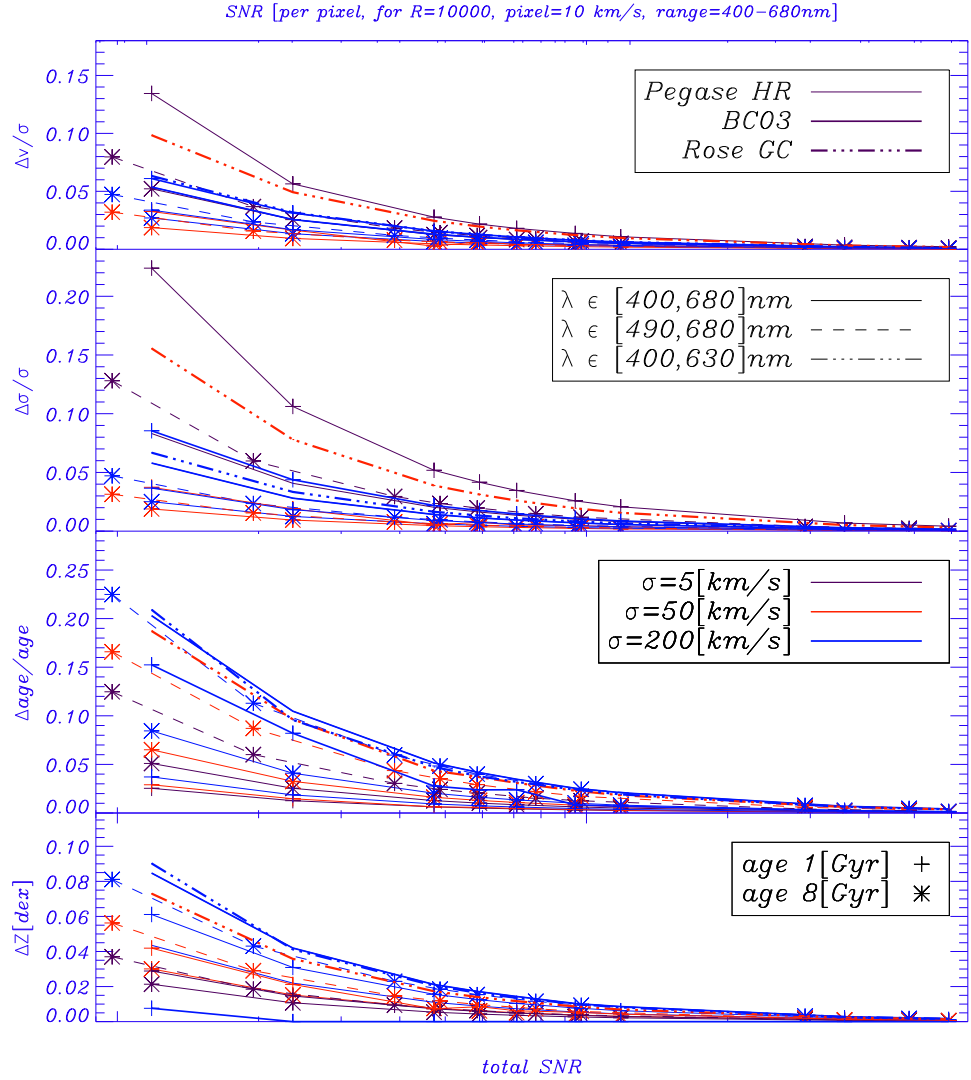}
  \caption{Error on population and kinematical parameters as a function of 
    the total SNR. The upper x axis shows SNR per pixel for $\lambda$ range
    from 400 to 680~nm and dispersion of 0.2\AA/pix.
    The curves noted 'BC03' are representing simulations
    with Bruzual \& Charlot (2003) models, and those noted 'Rose GC'
    are observation of globular clusters by Rose et al. (2005), all analyzed with
    PEGASE.HR templates. Errors are proportional to (total SNR)$^{-1}$, see the
    formula in the text.}
  \label{fig:2}
\end{figure}

\emph{Relation between precision and SNR.}
The precision essentially depends on the
total SNR integrated over the whole wavelength range 
(Fig~\ref{fig:2}): a lower SNR per
pixel can be balanced by a larger number of pixels. 
When the wavelength range is shortened to exclude
the blue region containing H$_\beta$ line and bluer (but keeping the same
total SNR), the precision on kinematics and metallicity is not seriously
affected, but the precision on age becomes twice worse.

\begin{figure}
  \centering
  \includegraphics[height=15cm]{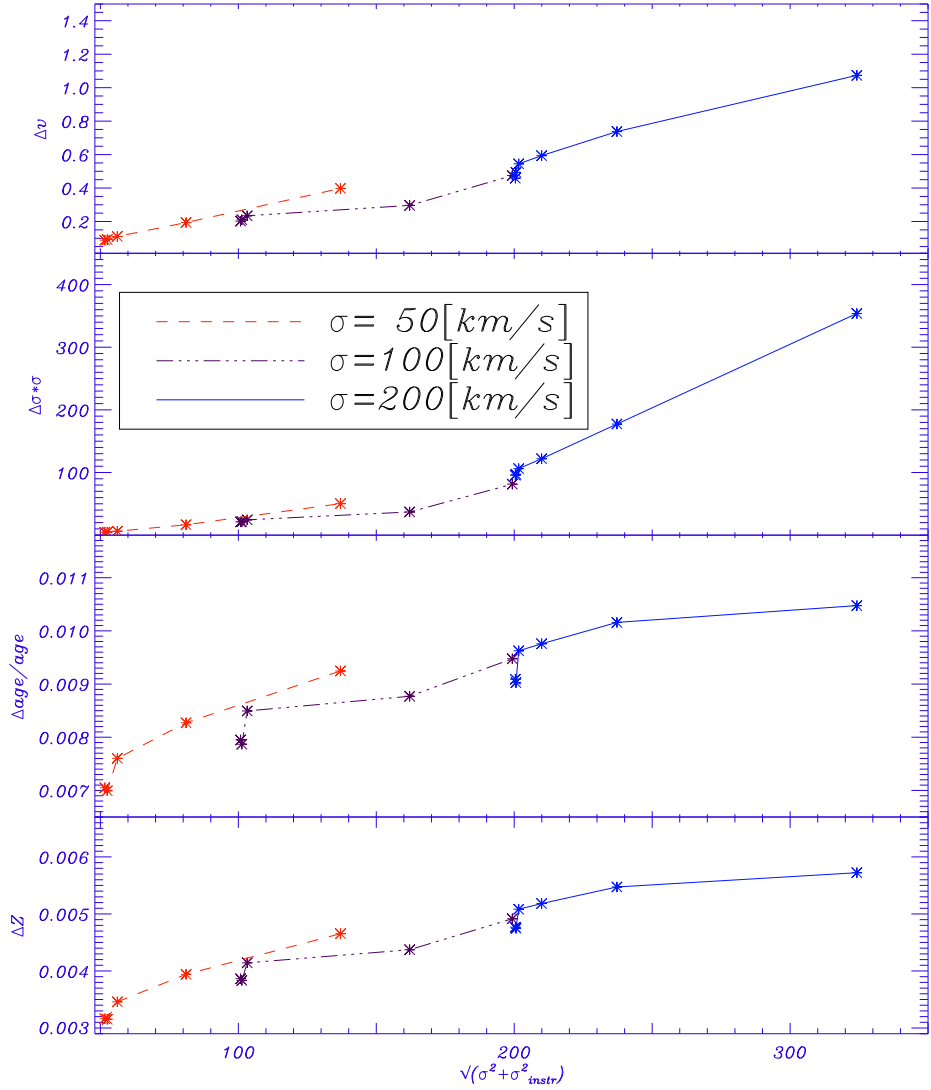}
  \caption{Errors on the population and kinematical parameters as a function of
    the resolution. The instrumental dispersion ranges from $\sigma_{ins}$ = 12~km/s
    to $\sigma_{ins}$ = 250~km/s (R=500 to 10000). Errors are increasing with the 
    effective broadening $\sqrt{(\sigma^2+~\sigma_{\mathrm{ins}}^2)}$ according to 
    the formula given in the text.}
  \label{fig:3}
\end{figure}

\emph {Relation between resolution and precision}. 
The precision on kinematical and population parameters
depends on ($\sigma^2+~\sigma_{\mathrm{ins}}^2$)(Fig~\ref{fig:3}), 
where $\sigma_\mathrm{ins}$ is the
instrumental velocity dispersion. The resolution is
an important parameter for the precise determination of the kinematics.
But it has weaker influence over
the errors on age and metallicity .(See the relations below.) 
When $\sigma < 0.4 \, \sigma_{\mathrm{ins}}$,
the different minimization strategies that we experimented generally become
unstable to measure the internal kinematics.

\section{Conclusion}
The method based on the pixel fitting with PEGASE.HR templates can
efficiently constrain kinematics and stellar population. We summarize below
the relations between the errors on the population characteristics, the total
signal to noise ratio, and the resolution (for the spectral range 4000\AA --
6800\AA):

\begin{center}
\begin{tabular}{rl}
$err_{\mathrm{v}}$ & $\approx5\times SNR^{-1}\times (\sigma^2+\sigma_{\mathrm{ins}}^2)^{2/3}$\\

$\sigma\times err_{\sigma}$ & $\approx 5\times SNR^{-1}\times (\sigma^2+\sigma_{\mathrm{ins}}^2)^{6/5}$\\

$err_{t}/t$ &$ \approx 54\times SNR^{-1}\times (\sigma^2+\sigma_{\mathrm{ins}}^2)^{1/14}$\\

$err_{Z}$ & $\approx 12\times SNR^{-1}\times (\sigma^2+\sigma_{\mathrm{ins}}^2)^{1/7}$\\
\end{tabular}
\end{center}

These relations are useful to select the observational setup: what is the
optimal compromise between the spectral dispersion and the size of the spatial
elements that minimizes the errors in a given observing time? Considering only
the sources of uncertainties modeled in our simulations, it appears 
that the best precision on the internal kinematics will be obtained
when $\sigma_{\mathrm{ins}} \approx 0.8\,\sigma$. For the best precision on 
the parameters of the stellar population it is preferable to maximize the
total SNR with a lower dispersion. Still, a single setup with a resolution
matching the velocity dispersion is in general the good choice.

The template mismatch due to abundance effects and the uncertainties
in the modeling of a stellar populations are probably the source
of significant biases on the parameters of the stellar population.
In particular, we can guess that if Balmer lines are not in the wavelength
range, the determination of age will be extremely sensitive to abundance
mismatch. We have inverted observations of globular clusters \cite{rose05} 
and simulated spectra from Bruzual \& Charlot
\cite{bc03}. For old populations the estimates appear too young and
metallic. We are investigating the origin of this problem that may
be connected with the bias found by Prieto et al. \cite{pri05} in the
ELODIE library.

\printindex
\end{document}